\documentclass[aip,apl,numerical,reprint,amsmath,amssymb]{revtex4-2}
\usepackage{graphicx}
\usepackage{dcolumn}
\usepackage{bm}
\usepackage{fullpage} 
\usepackage{amstext}
\usepackage{amsthm}
\usepackage{amssymb}
\usepackage{amsmath}
\usepackage{enumitem,booktabs}
\usepackage[referable]{threeparttablex}
\renewlist{tablenotes}{enumerate}{1} 
\makeatletter
\setlist[tablenotes]{label=\tnote{\alph*},ref=\alph*,itemsep=\z@,topsep=\z@skip,partopsep=\z@skip, 
parsep=\z@,itemindent=\z@,labelindent=\tabcolsep,labelsep=.2em,leftmargin=*,align=left,before={\footnotesize}} 
\makeatother                                           
\usepackage{multirow}
\usepackage{pifont}
\usepackage[normalem]{ulem}
\usepackage[usenames,dvipsnames]{color}
\usepackage{bbding}
\usepackage{pifont}
\usepackage{natbib}
\usepackage{xr}
\begin{document}

\title{Enhancement of polarization and magnetization in polycrystalline magnetoelectric composite}
\author{K.P. Jayachandran}
 \email{kpjayachandran@gmail.com}
\author{J.M. Guedes}
\author{H.C. Rodrigues}
\affiliation{IDMEC, Instituto Superior T{\'e}cnico, University of Lisbon, Av. Rovisco Pais, 1049-001 Lisbon , Portugal}
\date{\today}

\begin{abstract}

%
Electrical control of magnetization or magnetic control of polarization 
offers an extra degree of freedom in  materials possessing both electric 
and magnetic dipole moments \emph{viz.,} magnetoelectric multiferroics.  Microstructure
  with  polycrystalline configurations that enhances the overall 
  polarization/magnetization and that outperform single crystalline configurations are identified.
  The characterization of local fields corresponding to the polycrystal configuration 
  underlines nontrivial 
  role played by randomness in better cross-coupling mediated by anisotropic and asymmetric 
  strains. 
 
\end{abstract}

%
\maketitle
The magnetoelectric (ME) effect manifests in the linear relation between the magnetic and 
electric fields in matter and it causes, for instance, either a magnetization 
or electrical polarization proportional to applied electric/magnetic 
fields \cite{landau1960}. Yet, the ME effect is possible only for certain magnetic 
symmetry classes, akin to piezomagnetism.
Set of materials classified as magnetoelectric multiferroics  possess both the 
magnetic and ferroelectric orders in the same phase\cite{Schmid1994}. Apart from exhibiting 
 functionalities of both the orders, a coupling between the ferri-/ferro-magnetic (FM) 
 and ferroelectric (FE) states enable appearance of novel characteristics not present 
 in either of the states  \cite{Fiebig2016}. Nonetheless, simultaneous occurrence of 
 magnetism and ferroelectricity in  materials is constrained by the 
 conflicting classic chemical requirements regarding electronic orbital occupancy 
 \cite{Hill2000,Spaldin-Rev-NatMater-2019}. 
In such a situation, one should envisage only a moderate  ME
coupling though \cite{Mosotovoy-NatMater-2007}. For instance, only few examples 
exist of materials possessing intrinsic ME coupling at room temperature 
\cite{Mandal-Nature-2015, Long-Science-2020}. 
An unambiguous resolution to circumvent this chemical 'contraindication' is by juxtaposing 
an FE and FM material artificially 
into  strain-mediated multiphase materials such as composites combining the two phases, 
that thereby yields substantial ME effect 
\cite{Spaldin-Rev-NatMater-2019,Mosotovoy-NatMater-2007, harshe1994}. Moreover, 
the composites do not need to adhere to the symmetry restrictions of single phase 
materials and consequently do possess the freedom of choice from a wide variety of 
ferroics existing above room temperature. Crystallites of the two phases in an ME composite are 
assumed to be in good mechanical contact if the two phases are polycrystalline. 
The change in shape of the FE grains in response to an applied electric field causes 
the ferromagnetic grains to deform, resulting in a change in 
magnetization \cite{vandenBoomgaard1978}.
\begin{figure*}
	\centering
	\includegraphics [scale=0.75]{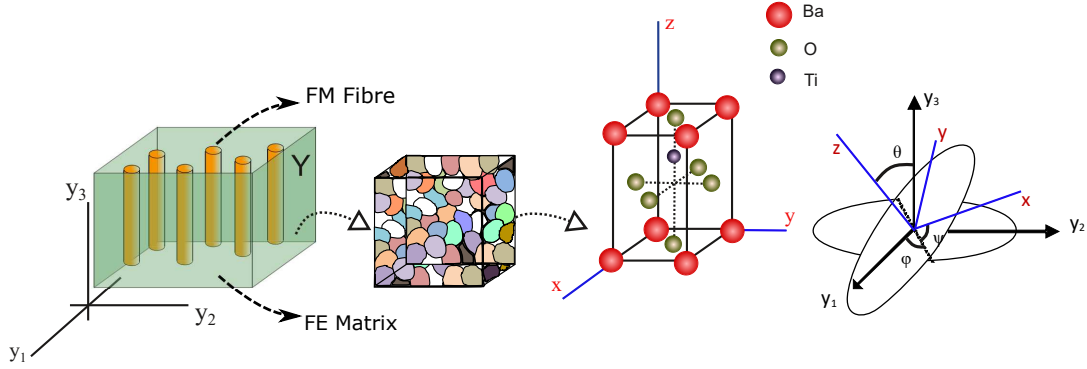}
	\caption{\label{figure:1}  Schematic hierarchical diagram of the ME multiferroic 
		microstructure browsing through the succession of modeling tasks in the study. 
		The microstructure $Y$ is a 
		fibre-reinforced ME composite where an FM fibre made of CoFe$_2$O$_4$ is embedded 
		in a perovskite FE matrix of BaTiO$_3$. BaTiO$_3$ is a polycrystal -- an aggregate of 
		grains of various orientations (depicted in different colours). The orientation of 
		the grains quantified by the Euler angles $(\phi,\theta,\psi)$ is in fact that of  
		the underlying 
		crystallographic unit cell and is measured against the microscopic ($y_1,y_2,y_3$)
		coordinate system.}
\end{figure*}

Magneto-elastic effect by means of strain transfer across the interface of an 
FM/nonmagnetic bilayer could unleash significant changes in the magnetic properties 
of the ferromagnet. When an FE material is used as the nonmagnetic layer, 
strain transfer can be reinforced through the induction by an
electric field thanks to the inverse piezoelectric effect prevails in the FE material. 
This additional strain transfer
allows one to manipulate the magnetisation by an electric
field \cite{Taniyama-JPC-2015}. The concept of magnetic information recording and
memory device technologies as well as electric-field driven magnetoelastic 
effects on magnetic anisotropy have attracted much
interest recently from both fundamental and technological
aspects owing to the said modification accomplished on magnetisation orientation 
\cite{nan2008,Taniyama-JPC-2015,Spaldin-Rev-NatMater-2019}. Among  many
FM/FE hybrid structures, perovskite manganites/BTO are the
well-suited  magnetoelastic  heterostructures  since  transport,
magnetic, and electronic properties of perovskite manganites
are very sensitive to lattice strain and can be easily integrated
with  BaTiO$_3$ (BTO)  due  to  the  similar  crystal  
structure \cite{Panchal-PhysRevB-2018}.  For instance 
when an electric field is applied to the nanostructured BTO/CFO 1--3 
composite, the strain
created by the piezoelectric (BTO) matrix couples to the magnetostrictive 
CFO(CoFe$_2$O$_4$) pillars and hence the strength of the coupling between the 
ferroelectricity and magnetism are critically dependent on the  anisotropy 
and crystalline structure of the constituents\cite{ZhengRamesh2004, HaimeiZheng2007}.

In this letter, we would be studying the impact of an applied electrical/magnetic 
fields on a motley ME composite composed of disparate FE and FM materials. We 
underscore the significance of heterogeneity of  one of the constituent 
phases, \emph{viz.,} the FE phase by introducing randomness in the microstructure.  
Prior research have shown that introducing 
heterogeneity locally can improve the piezoelectric properties of FE materials 
and the ensuing coupling effect of the ME composite of which the former is 
a component \cite{Chen-Shrout-NatMater-2018,jayachandran2008}. 
Apart from the effect of applied fields, the sway of crystal anisotropy
on the polarization/magnetization as demonstrated in various studies 
\cite{Sahoo-PRB-2007,Han-APL-2014,Zhang-SciRep-2014, Zhou-SciRep-2014}on 
magnetoelectric heterostructures is also shown. 
 Another important 
challenge is to identify the optimum texture and stress level and efficient
control of the same to achieve targeted magnetoelectric 
properties \cite{Stephen-NatMater-2019}. We demonstrate the effect of local texture 
in a 1--3 composite, tantamount to the architecture used in Zheng et al., 
\cite{ZhengRamesh2004} to map its local field distributions upon applying an 
external bias field. Moreover, the 
average polarization and magnetization post application of ample bias fields 
sufficient to saturate the composite would be computed.

Estimation of the equilibrium macroscopic magneto-electro-electric properties of the ME 
composite,have been done using the mathematical homogenization method
\cite{sanchez1980,Jayachandran-ScRep-2020}.  A ME multiferroic composite occupying a 
volume $\Omega$ of coordinates $\mathbf{x}$ (or $x_i$) is considered 
and the
material properties are assumed to change periodically and that period $\epsilon$ 
is characterized by the dimension of an
elementary cell $\mathrm{Y}$ of coordinates $\mathbf{y}$ (or $y_i$) of the body 
(Fig.~\ref{figure:1}). This lead us to 
examine fields for unknown physical quantities in the form of two-scale asymptotic 
expansion in powers of $\epsilon$ as
\begin{equation}
\label{eq:1}
\zeta^{\epsilon}(\mathbf{x})=\zeta^{0}(\mathbf{x,y})+\epsilon
\zeta^1(\mathbf{x,y}),\text{with}~\mathbf{y}=\mathbf{x}/\epsilon\\
\end{equation} 
and the equivalent macroscopic behaviour is estimated to be first order by the behaviour of 
 $\zeta^{0}$. Here $\zeta$ stands in lieu of the fields \emph{viz.,} 
 $\mathbf{u}$, the displacement, $\varphi$ and $\psi$, the scalar electrical/magnetic 
 potentials respectively of a multiferroic composite \cite{Jaya2014}. Here the functions 
$\zeta^1(\equiv \mathbf{u}^1,\varphi^1~ \text{or}~ \psi^1)$ are the local variations (or 
\emph{field perturbations}) describing the heterogeneous part of the solutions and are 
associated with $\mathbf{y}=\mathbf{x}/\epsilon$. Applying calculus of variations, 
utilising the asymptotic Eq.~(\ref{eq:1}) and its derivatives, the homogenization 
method essentially culminates in the characterization of effective magneto-electro-elastic 
moduli when the characteristic length $\epsilon$ of the period tends to zero. The electrical, 
magnetic and mechanical fields (\emph{degrees of freedom})  govern the constitutive 
equations of the linear ME multiferroic solid (see Supplementary material). The 
model quantifies all the local fields and potentials besides stress ($\mathbf{\sigma}$) 
 and strain ($\mathbf{\varepsilon}$) fields, 
magnetization (via magnetic flux density, $\mathbf{B}$), polarization 
(via electric displacement, $ \mathbf{D}$ as well as von-Mises stress upon 
the application of an external field (either electric/magnetic or mechanical). 
The \emph{macroscopic averages} $\langle\sigma_{ij}\rangle$, $\langle D_{i}\rangle$, and 
$\langle B_{i}\rangle$  can be computed
once the homogenized solution for $\mathbf{u}^0$, $\varphi^0$ and $\psi^0$ and that of the 
corresponding fields \emph{viz.,} $\frac{\partial u^0_j(\mathbf {x})}{\partial x_k}$, 
$\frac{\partial \varphi^0(\mathbf {x})}{\partial x_j}$ and 
$\frac{\partial \psi^0(\mathbf {x})} {\partial x_j}$ are prescribed (see Supplementary 
material). 
This \emph{postulate} 
is equally applicable for the cases   
when magnetic field $\textbf{H}=0$, the applied macroscopic electric field 
$\langle E_{i}\rangle$ generates average magnetization \cite{landau1960}
\begin{equation}
\label{eq:2}
\langle M_{k}\rangle=\widetilde{\alpha}_{ik}\langle E_{i}\rangle
\end{equation}
and when  electric field $\textbf{E}=0$, the applied macroscopic magnetic field 
$\langle H_{i}\rangle$ generates average electrical polarization \cite{landau1960}
\begin{equation}
\label{eq:3}
\langle P_{i}\rangle=\widetilde{\alpha}_{ik}\langle H_{k}\rangle
\end{equation}
\begin{figure*}
	\centering
	\includegraphics [scale=0.85]{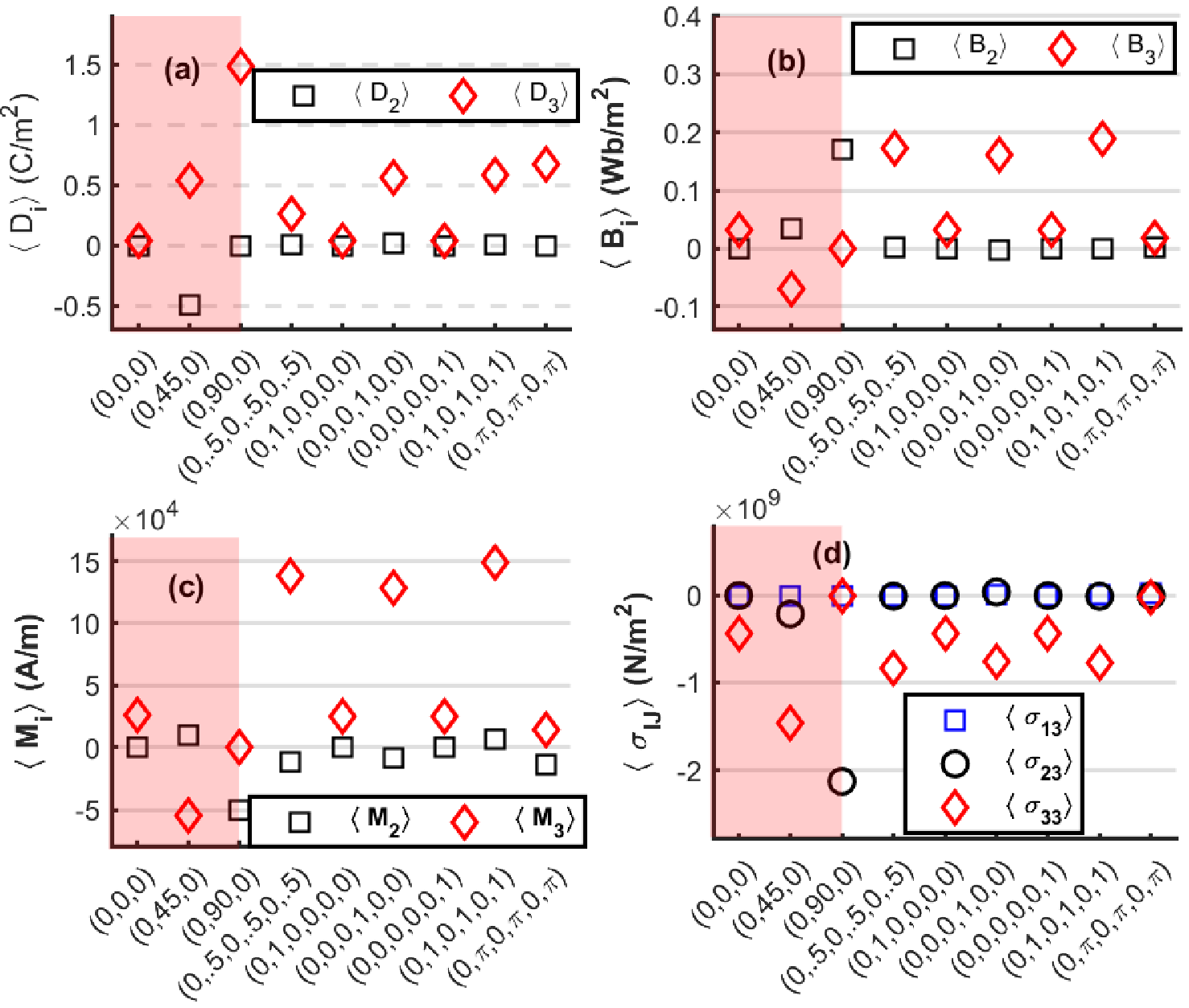}
	\caption{\label{figure:2}  Average fields for the multiferroic ME 1--3 composite 
		BTO--CFO. (a) shows the average dielectric displacement $\langle D_i \rangle$.
		(b) is the average magnetic flux density, $\langle B_i \rangle$ (c) shows the 
		average magnetization $\langle M_i \rangle$ and (d) show the relevant tensor components 
		of the average stress  $\langle \sigma_{IJ} \rangle$ upon applying an 
		external electric field $\langle E_3 \rangle$ along the $y_3$--axis of the 
		composite (i.e., along the normal to the composite plane). The data fall in 
		the shaded area corresponds to single crystal BTO matrix--polycrystalline 
		CFO fibre constituting the ME composite. The BTO orientations in Euler angles 
		$(\phi^\circ,\theta^\circ,\psi^\circ)$ are marked along the horizontal axis in the 
		shaded area. 
		Rest of the data correspond to polycrystalline ME composite with BTO orientation 
		distribution $(\mu_\phi,\sigma_\phi,\mu_\theta,\sigma_\theta,\mu_\psi,\sigma_\psi)$ 
		(along the horizontal axis).
		Here the x--axes is representative line separating the quantities plotted on 
		the y--axes. i.e., the data points in the plot have no abscissae.}
\end{figure*}
\begin{figure*}
	\centering
	\includegraphics [scale=0.70]{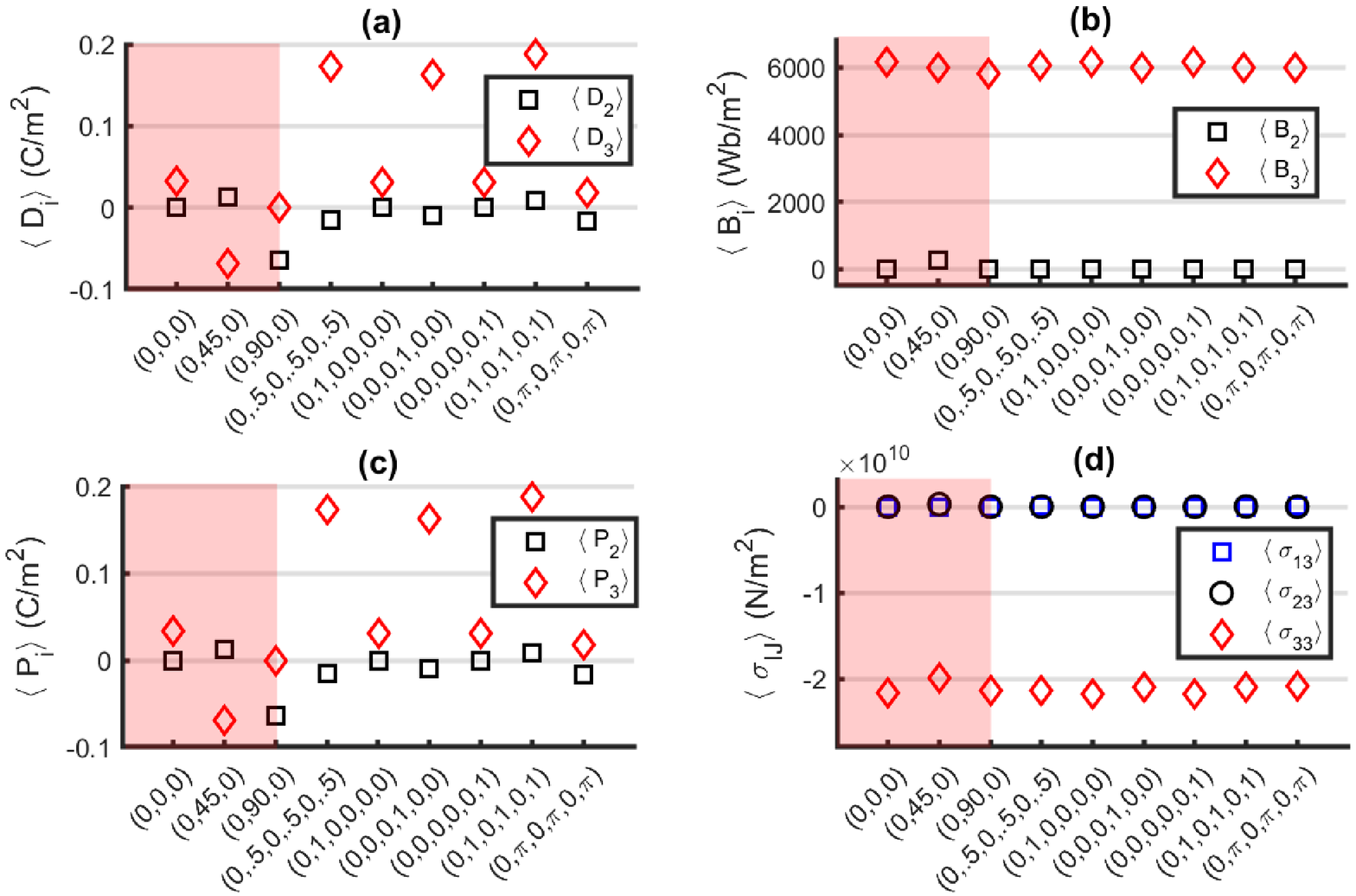}
	\caption{\label{figure:3}  Average fields for the multiferroic ME 1--3 composite 
		BTO--CFO. (a) shows the average dielectric displacement $\langle D_i \rangle$.
		(b) is the average magnetic flux density, $\langle B_3 \rangle$ (c) shows the 
		average magnetization $\langle M_i \rangle$ and (d) show the relevant tensor components 
		of the average stress  $\langle \sigma_{IJ} \rangle$ upon applying an 
		external magnetic field $\langle H_3 \rangle$ along the $y_3$--axis of the 
		composite (i.e., along the normal to the composite plane).
	Here the x--axes is representative line separating the quantities plotted on 
	the y--axes. i.e., the data points in the plot have no abscissae.}
\end{figure*}
\begin{figure}
\centering\includegraphics[scale=0.25]{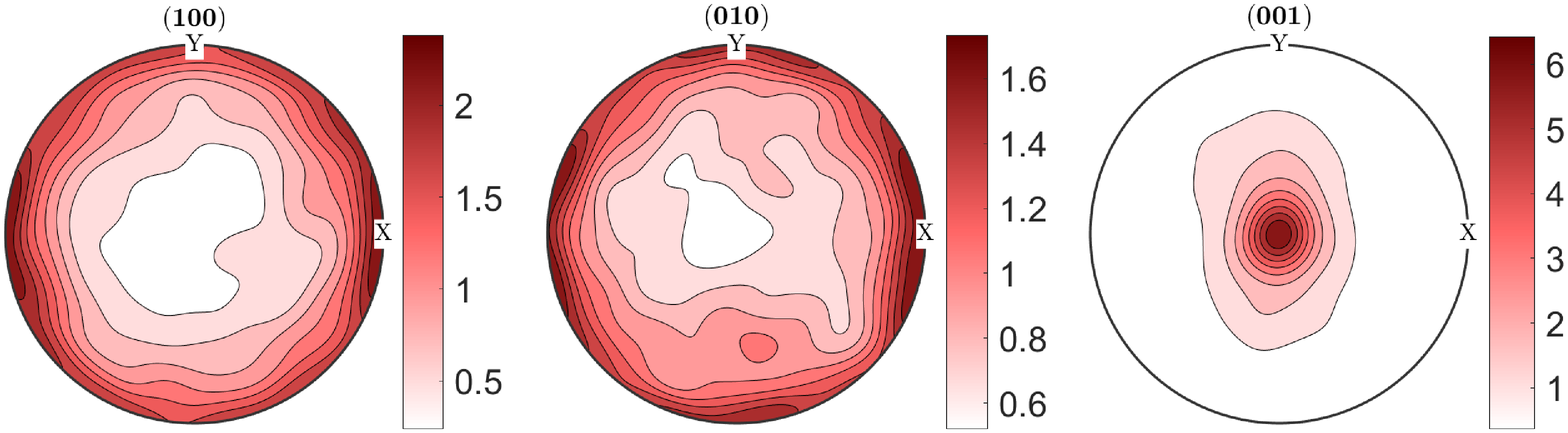}
\caption{\label{figure:4}Pole figures (contour plots) of the polycrystalline BTO matrix while the ODPs are 
$(\mu_\phi=0,\sigma_\phi=1,\mu_\theta=0,\sigma_\theta=1,\mu_\psi=0,\sigma_\psi=1)$. 
At this configuration, the ME composite deliver maximum $\langle M_3 \rangle$ and 
$\langle P_3 \rangle$.}
\end{figure}

The FE matrix of the ME composite is composed of polycrystalline BaTiO$_3$. The orientation 
of the BaTiO$_3$ crystallographic grains in the
composite is characterized using Euler angles. It is seen from the Euler angles 
$(\phi, \theta,\psi)$ (Fig.~\ref{figure:1}) that it measure the rotation of the 
crystallographic coordinates (x, y, z) with reference
to the microstructure coordinates ($y_1, y_2, y_3$). The three Euler angles are 
modelled as
statistically distributed in a normal distribution after poling
obeying the probability distribution function \(  
f(\alpha\mid\mu,\sigma)=1/(\sigma\sqrt{2\pi})
exp-[(\alpha-\mu)^2/2\sigma^2]\). Here $\alpha$ is the random variable representing 
 the orientations (i.e., the Euler angles) $(\phi, \theta,\psi)$ and $\mu$ and 
$\sigma$  are the  mean and the standard deviation, respectively.  The fibres embedded 
 in the 
matrix is composed of the FM material CoFe$_2$O$_4$ which is treated as bulk 
material without assigning any orientation whatsoever. The numerical model developed 
is implemented in finite element method. The convergence of piezoelectric properties 
with unit cell size allows us to determine the simulation-space independent, 
macroscopic magnetoelectric properties at various
distribution of grains (See Supplementary material). 

\begin{table}
	\caption{\label{tab:table1}Average values of magnetization $\langle M_{3}\rangle$ (in A/m) at 
		constant external electric field, and electric polarization $\langle P_{3}\rangle$ (C/m$^2$) 
		at external magnetic 
		field of 1–3 magnetoelectric 
		composite BaTiO$_3$ (single and poly-crystalline)–ceramic CoFe$_2$O$_4$. }
	\begin{ruledtabular}
		\begin{tabular}{lccr}
			BTO phase&Orientation \footnote{Euler angles $(\phi,\theta,\psi)$ (rad) of rotation in single 
				crystal BTO phase} &\multicolumn{2}{c}{Averages}\\
			&or ODP\footnote{Orientation Distribution Parameters $(\mu_\phi, \sigma_\phi)\|%
				(\mu_\theta,\sigma_\theta)\|(\mu_\psi,\sigma_\psi)$ (rad) of polycrystal BTO phase} 
			&$\langle M_{3}\rangle$&$\langle P_{3}\rangle$\\
			\noalign{\smallskip}
			\hline
			\multirow {3}*{Single crystalline}&(0,0,0) &2.63$\times10^4$ &0.03\\
			&(0,$\pi/4$,0)&-5.48$\times10^4$&-0.07\\
			&(0,$\pi/2$,0)&-1.39            &-2$\times10^{-6}$\\
			\noalign{\smallskip}
			\hline
			\noalign{\smallskip}
			\multirow {7}*{Polycrystalline}   &(0,$\frac{1}{2}$)$\|$(0,$\frac{1}{2}$)$\|$(0,$\frac{1}{2}$)&1.38$\times10^5$&0.17\\
			&(0,1)$\|$(0,0)$\|$(0,0)      &2.47$\times10^4$&0.03\\
			&(0,0)$\|$(0,1)$\|$(0,0)      &1.3$\times10^5$ &0.16\\
			&(0,0)$\|$(0,0)$\|$(0,1)      &2.47$\times10^4$&0.03\\
			&(0,1)$\|$(0,1)$\|$(0,1)      &1.49$\times10^5$&0.19\\
			&(0,$\pi$)$\|$(0,$\pi$)$\|$(0,$\pi$)&1.46$\times10^4$&0.02\\			                            
			Experiment \footnote{CoFe$_2$O$_4$ nanopillars embedded BaTiO$_3$ matrix in a 1--3 composite  
				from Ref.~[\onlinecite{ZhengRamesh2004}].}&--& 
			$\approx$ 3.5$\times10^5$&$\approx$ 0.23\\
			Experiment \footnote{CoFe$_2$O$_4$--BaTiO$_3$ 1--3 nanocomposite  
				from Ref.~[\onlinecite{Carolin-NatCom-2013}].}&--& 
			$\approx$ 3.15$\times10^5$&$\approx$ 0.1\\
		\end{tabular}
	\end{ruledtabular}
\end{table}

The average magnetic property response along the $y_3$--axis of the local 
coordinates (which was set to align along the normal to the composite plane) consequent
to an external biasing electrical field is studied first. Here we apply an electric 
field ( $E_3$ ) which suffices to saturate the polycrystalline BTO--CFO composite. The 
results are summarised in Fig.~\ref{figure:2}. In general, the average fields acting 
through out-of-plane (the components along $y_3$--axis)  to the composite plane are 
greater in magnitude than the other components. Unlike the common perception of great performance 
of single crystal piezoelectrics, 
we see much better magnetization $\langle M_{3}\rangle$ while 
the BTO is still polycrystalline (Fig.~\ref{figure:2} and Table~\ref{tab:table1}). 
In Fig.~\ref{figure:2} the polycrystalline data of the averages are differentiated from single 
crystal by painting the single crystal region by a shade. Randomness, introduced by local 
microstructural heterogeneity 
could potentially enhance piezoelectricity in relaxor ferroelectric ceramics 
\cite{Chen-Shrout-NatMater-2018}. Here, we have incorporated randomness in the 
ferroelectric BTO phase through Euler orientations of the constituent grains kept 
at a normal distribution but with a standard deviation of $\mu$ rad. The maximum 
magnetization  $\langle M_{3}\rangle$  is seen for BTO polycrystal phase with 
orientation distribution parameters $(\mu_\phi=0,\sigma_\phi=1,\mu_\theta=0,
\sigma_\theta=1,\mu_\psi=0,\sigma_\psi=1)$. Each grain 
orientation would be dissected and kept separately and the grains are allowed 
to chose the combination of three  $(\phi,\theta,\psi)$ but dictated by the 
corresponding mean ($\mu$) and standard deviation ($\mu$). The value of 
$\langle M_{3}\rangle =1.49\times10^5 (A/m)$ obtained here compares with the order 
of $\langle M_{3}\rangle$ measured by Zheng et al., \cite{ZhengRamesh2004} on 
nanostructures with vertically aligned CoFe$_2$O$_4$ nanopillars embedded in 
BaTiO$_3$ matrix. Zheng et al. value is the saturation magnetization ($M_s$) response 
against an applied magnetic field (hysteresis) of the composite rather than the 
cross coupling magnetization value as is obtained in the present work (Table~\ref{tab:table1}) 
and hence the deviation is obvious. 
Epitaxial CFO films on BaTiO$_3$ single crystal substrate shows  $\langle M_{3}\rangle
  \approx 2.5\times10^5~(A/m)$ \cite{Chopdekar-APL-2006} under a magnetic field.

The property variation at various configurations (single crystalline data in shaded area) 
consequent to the application of external magnetic field $H_3$ are given in Fig.~\ref{figure:3}.
Here the electric field is off and hence the electrical responses are dictated solely by the 
piezomagnetic effect of the CFO component of the ME composite. More precisely, the 
emergence of electrical displacement $\langle D_j \rangle$ (Fig.~\ref{figure:3}) which 
bears the signature of polarization and the electrical polarization 
$\langle P_j \rangle$ (Fig.~\ref{figure:3} and Table~\ref{tab:table1}) itself. The average 
polarization out-of-plane of the composite layer $\langle P_3 \rangle$ peaks at the 
polycrystal phase of BTO as seen in magnetization $\langle M_3 \rangle$ in the previous 
simulation experiment with applied electric field (Fig.~\ref{figure:2}).   The polarization 
$\langle P_3 \rangle$ compares well with the experiments (Table~\ref{tab:table1})

The causal relationship between  local microstructure and better magnetic response of polycrystalline 
BTO--CFO composite can be explored from  mapping the local field distribution in response to the external field. 
The average or macroscopic field applied to a 
magnetoelectric composite would permeate the 
material microstructure and would spread unequally into different points owing to the 
heterogeneity of the material. The anisotropy due to the underlying crystalline structure  
of the constituent phases contribute to this phenomenon. The associated local field 
distribution exhibit spacial fluctuations critical to coupling phenomena. The stress/strain 
mediating the coupling phenomenon is spread non-uniformly across the 
microstructure as is seen (in Fig.~S\ref{fig:S1}) in Supplementary material. The equivalent 
von Mises stress indicates a stress concentration across BTO matrix compared to the 
CFO fibre. This is reflected and underlined in the variation of microscopic displacement 	
$\mathbf{u}^{\varepsilon}$ (see Fig.~S\ref{fig:S1}). The asymmetry and anisotropy of the local 
stress/strain 
distributed across the microstructure could be seen in  histograms (Fig.~S\ref{fig:S2}). 
 The in-plane strain $\epsilon_{12}^\varepsilon$ records values orders of magnitude greater
 than other components (Fig.~S\ref{fig:S2}) conforming the compression of the 
 composite in $x$ and $y$ directions \cite{Carolin-NatCom-2013}. The cross-coupling 
 effect of the applied electric field results in appreciable local magnetic potential 
 $\psi^\varepsilon$ and associated magnetism (Fig.~S\ref{fig:S3}). The magnetic field
 induced local profile of fields are shown in the Supplementary materials. It reinforces 
 the cross-coupling between the magnetic and electric degrees of freedom through 
 mechanical stress. The pole figure in Fig.~\ref{figure:4} shows the distribution of 
 grain orientation about the $y_3$--axis (or $z$--axis) of the composite structure.
 The $<001>$ axes of the ferroelectric BTO (where the spontaneous polarization is oriented ) 
 is aligned mostly along the   $y_3$--axis (here the $[001]$ psuedo cubic axes of BTO 
 is directed ou-of-plane). 
 
 In summary, strong magnetoelectric coupling is resulted while a polycrystalline 
 ME composite of  1--3 BaTiO$_3$--CoFe$_2$O$_4$ is subjected external electric/magnetic 
 fields. In contrast to single crystal BaTiO$_3$--polycrystal CoFe$_2$O$_4$ composite, 
 the averages of polarization and magnetization of the polycrystalline 
 BaTiO$_3$--CoFe$_2$O$_4$ exceeds that of the single crystal version. The depiction 
 of local fields corresponding to the polycrystal configuration suggests nontrivial 
 role played by randomness in better cross-coupling mediated by anisotropic and asymmetric 
 strains.

\section{Acknowledgments}
This work was supported by  FCT-Funda\c{c}\~{a}o para a Ci\^{e}ncia e a Tecnologia, 
through IDMEC, under LAETA, project UIDB/50022/2020.
%
%
\end{document}


\renewcommand{\theequation}{S\arabic{equation}} 
\title{Supplementary material to 'Enhancement of polarization and magnetization 
	in polycrystalline magnetoelectric composite'}
\author{K.P. Jayachandran}
\email{kpjayachandran@gmail.com}
\author{J.M. Guedes}
\author{H.C. Rodrigues}
\affiliation{IDMEC, Instituto Superior T{\'e}cnico, Universidade de Lisboa, 
	Lisboa, Av. Rovisco Pais, 1049-001 Lisbon , Portugal}
\begin{abstract}
This supporting material for the work on 
microscopic characterization of the composite polycrystalline magnetoelectric 
multiferroics include 
Literature review, theory of homogenization, its numerical implementation and few key 
results.
\end{abstract}
\maketitle                    
\section{Introduction}
The materials classified as magnetoelectric multiferroics  possess both the magnetic and 
ferroelectric orders in one phase\citep{Schmid1994}. Substantial ME effect can be derived 
through  fabricating 
composites of a ferroelectric (FE) and a ferromagnetic (FM) material in the form 
composites \cite{harshe1994}.
We have used homogenization theory 
for the analysis of the external field (electrical or magnetic) applied on a magnetoelectric 
composite and the resultant polarization and magnetization responses which are vital in its 
use as new devices. When the period of the 
structure is very small, a direct numerical approximation of the solution
to magnetoelectric problem may be prohibitive, or even impossible. Here homogenization 
provides an alternative scheme of
approximating such solutions by means of a function which solves the problem 
corresponding to a “homogenized” material. Homogenization method deals with the asymptotic 
analysis of partial differential equations in heterogeneous materials with a periodic 
structure, when the characteristic length $\epsilon$ of the
period tends to zero.
 \begin{figure*}
	\centering
	\includegraphics [scale=0.28]{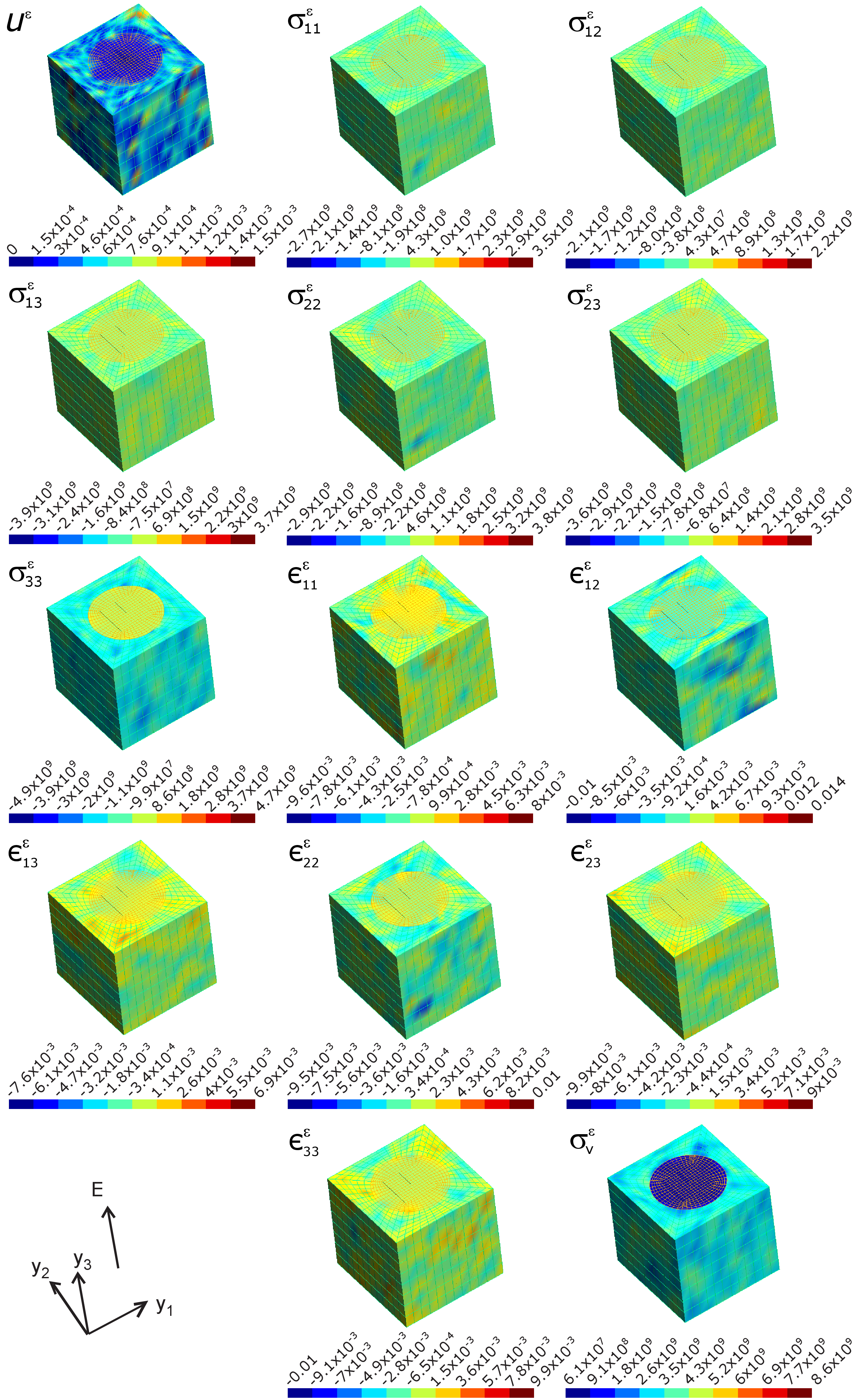}
	\caption{\label{fig:S1}  Map of local fields (computed at the nodal points of the FEM)
			of magnetoelectric composite BaTiO$_3$--CoFe$_2$O$_4$, 
			\emph{viz.}, the stress $\sigma_{ij}^{\varepsilon}~(N/m^2)$ ,
			strain $\epsilon_{ij}^{\varepsilon}$, the equivalent von Mises stress 
			$\sigma_{v}^{\varepsilon}~(N/m^2)$, and displacement  
			$\mathbf{u}^{\varepsilon}~(m)$ upon 
			applying a global electric field $\mathbb{E}$ of $10^8 V/m$ on the unit cell. 
			Here the magnetic CoFe$_2$O$_4$ cylindrical pillars are 
			surrounded by ferroelectric BaTiO$_3$ matrix and both are aligned towards 
			the y$_3$ axis of the local coordinate system.}
	\end{figure*}
\begin{figure*}
	\centering\includegraphics[scale=0.25]{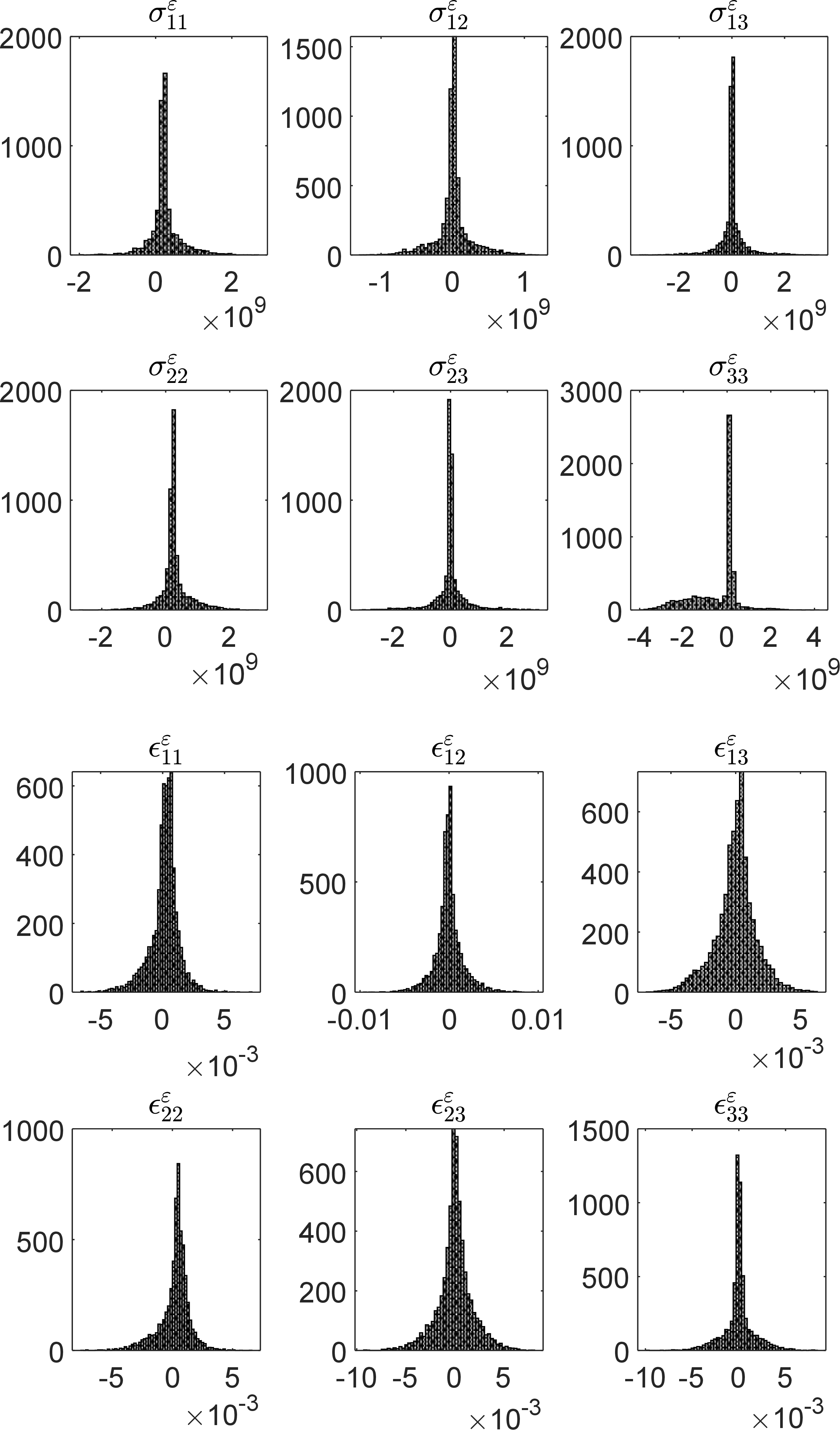}
	\caption{\label{fig:S2}Histogram of the local stress ($\sigma_{ij}^\varepsilon$) and 
		strain ($\epsilon_{ij}^\varepsilon$) values consequent to the application of external electric field (averaged over finite elements).}
\end{figure*}

 \begin{figure*}
	\centering
	\includegraphics [scale=0.28]{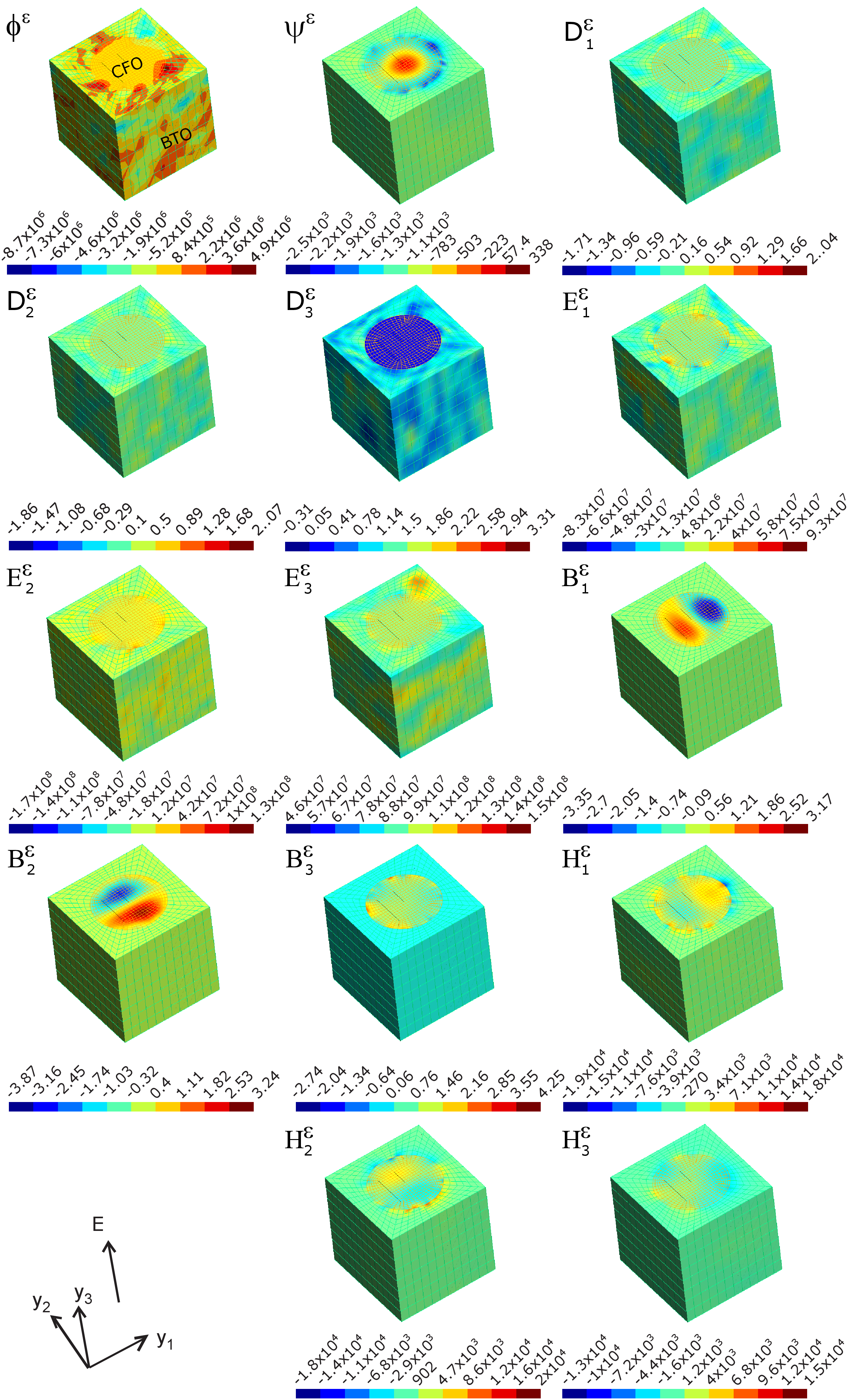}
	\caption{\label{fig:S3} Map of local  magnetic scalar potential $\psi^\varepsilon~(A)$,
		electric potential $\varphi^\varepsilon~(V)$, electric field 
		$\mathrm{E}_j^{\varepsilon}~(V/m)$, magnetic field $\mathrm{H}_j^{\varepsilon}~(A/m)$, 
		electric displacement $\mathrm{D}_3^{\varepsilon} (C/m^2)$, and magnetic flux
		$\mathrm{B}_3^{\varepsilon}~(Wb/m^2)$ computed at the nodal points of the FEM, 
		upon applying a global electric field $\mathbb{E}$ of $10^8 V/m$ on a 
		magnetoelectric composite of BaTiO$_3$--CoFe$_2$O$_4$.}
\end{figure*}
\begin{figure*}
	\centering
	\includegraphics[scale=0.28]{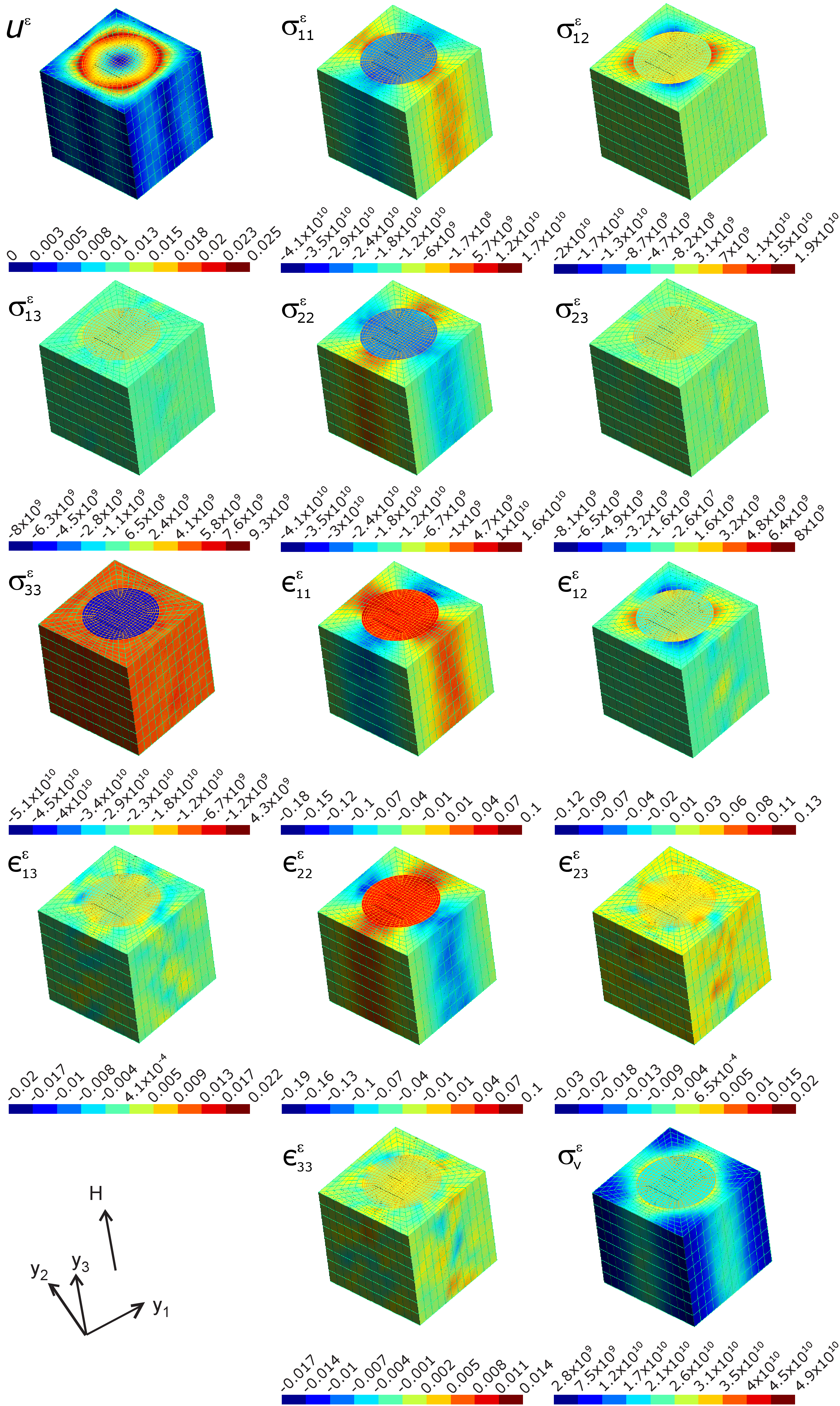}
	\caption{\label{fig:S4}  Map of local fields (computed at the nodal points of the FEM)
		of magnetoelectric composite BaTiO$_3$--CoFe$_2$O$_4$, 
		\emph{viz.}, the stress $\sigma_{ij}^{\varepsilon}~(N/m^2)$ ,
		strain $\epsilon_{ij}^{\varepsilon}$, the equivalent von Mises stress 
		$\sigma_{v}^{\varepsilon}~(N/m^2)$, and displacement  
		$\mathbf{u}^{\varepsilon}~(m)$ upon 
		applying a biasing magnetic field $\mathbb{H}$ of $10^8 A/m$ on the unit cell. 
		Here the magnetic CoFe$_2$O$_4$ cylindrical pillars are 
		surrounded by ferroelectric BaTiO$_3$ matrix and both are aligned towards 
		the y$_3$ axis of the local coordinate system.}
\end{figure*}
\begin{figure*}
	\includegraphics[scale=0.3]{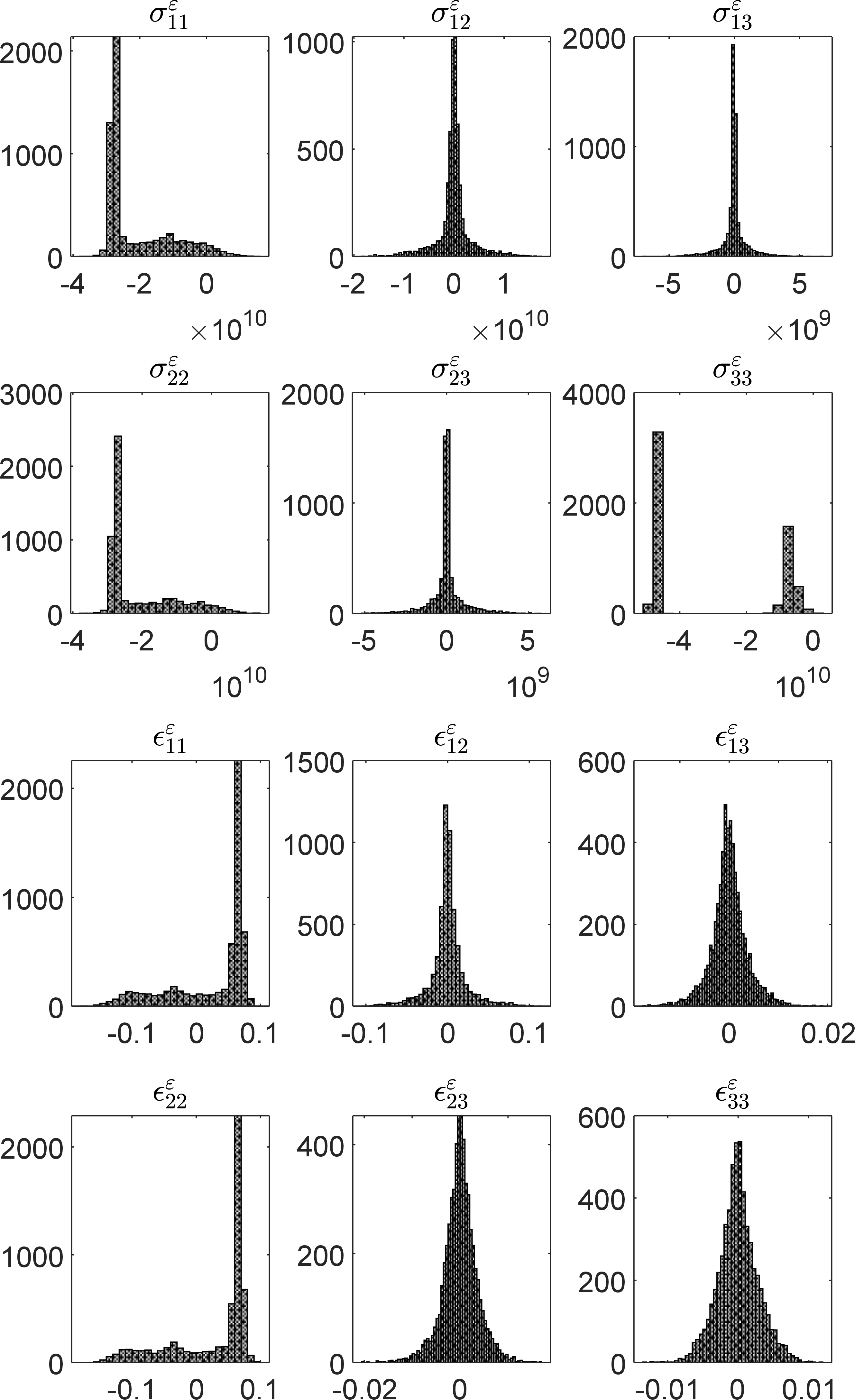}
	\centering
	\caption{\label{fig:S5} Histogram of the local stress ($\sigma_{ij}^\varepsilon$) and strain ($\epsilon_{ij}^\varepsilon$) values consequent to the application of external magnetic field (averaged over finite elements).}
\end{figure*}
\begin{figure*}
	\centering
	\includegraphics [scale=0.28]{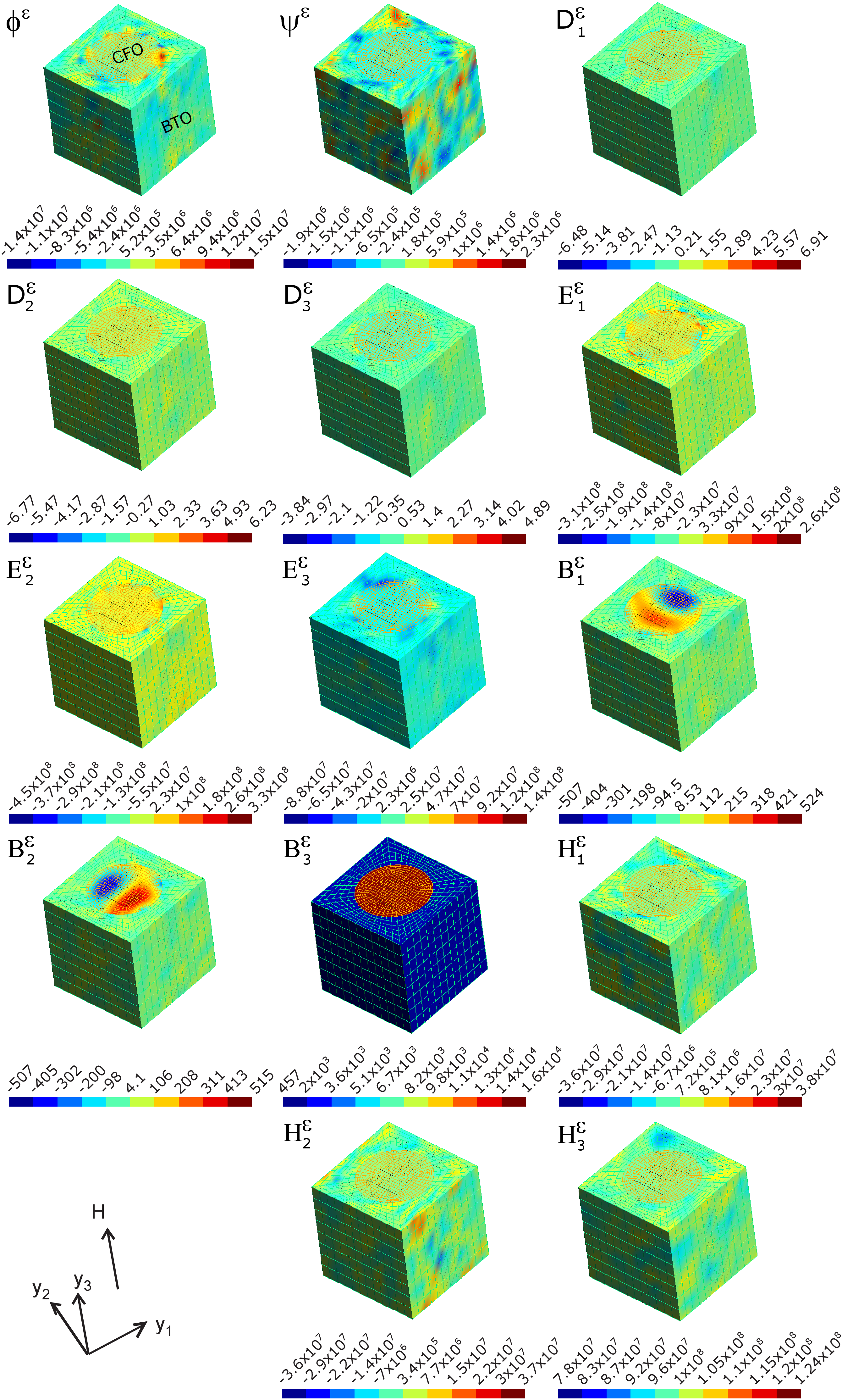}
	\caption{\label{fig:S6} Map of local  electric potential $\varphi^\varepsilon~(V)$, 
		magnetic scalar potential $\psi^\varepsilon~(A)$, electric field 
		$\mathrm{E}_j^{\varepsilon}~(V/m)$, magnetic field $\mathrm{H}_j^{\varepsilon}~(A/m)$, 
		electric displacement $\mathrm{D}_3^{\varepsilon} (C/m^2)$, and magnetic flux
		$\mathrm{B}_3^{\varepsilon}~(Wb/m^2)$ computed at the nodal points of the FEM, 
		upon applying a biasing magnetic field $\mathbb{H}$ of $10^8 A/m$ on a 
		magnetoelectric composite of BaTiO$_3$--CoFe$_2$O$_4$.}
\end{figure*}
 
\section{Theory}
 \subsection{Asymptotic homogenization}

A two-scale asymptotic homogenization analysis combined with a variational 
formulation is developed  for determination 
of the equivalent material properties of a periodic multiferroic magnetoelectric composite. 
Local and average (global) electrical, magnetic, and mechanical constitutive 
behaviour are computed.
 For a linear magneto-electro-elastic solid,
constitutive equations are governed by the electrical, mechanical
and magnetic fields. The  model quantifies the local electrical and magnetic 
potential, displacements,  electrical and 
magnetic fields,
stress and strain fields, magnetization (through magnetic flux density) and 
polarization (through electric displacement) and von-Mises
stress, besides the effective magneto-electro-mechanical properties.   
 The mathematical theory of homogenization accommodates the phase 
interaction in characterising both the macro- and micro-mechanical behaviours of the composite
material. i.e., the method permits the introduction of different field equations 
in a microscopic scale to each constituent of a composite while following the representative
volume element (RVE) notion.    
 Let $\mathbf\Omega$ be a fixed domain in $\mathbf{x}$--space. We consider
an auxiliary $\mathbf{y}$--space divided into parallelepiped periods $\mathbf{Y}$. 
The linear constitutive relations for small deformations for  multiferroics 
in the absence of heat flux are given by
\begin{eqnarray}
\label{eq:S8}
\sigma_{ij}&=& C_{ijkl}^{EH}\epsilon_{kl}-e_{kij}E_k-e^{M}_{kij}H_k 
\\
\label{eq:S9}
D_{i}&=& e_{ijk}\epsilon_{jk}+\kappa_{ij}^{\epsilon H}E_j+\alpha_{ij}H_j
\\
\label{eq:S10}
B_{i}&=& e_{ijk}^{M}\epsilon_{jk}+\alpha_{ji}E_j+\mu^{\epsilon E}_{ij}H_j
\end{eqnarray} 
Here  $\mathbf{\sigma}$, $\mathbf{\epsilon}$,  $ \mathbf{D}$, 
$\mathbf{B}$ are stress, strain, displacement, body force, mass density, electric displacement 
vector, and magnetic flux density respectively. $\mathbf{C^{EH}}$, $\mathbf{e}$, $\mathbf{e^M}$, 
$\mathbf{\kappa^{\epsilon H}}$ and $\mathbf{\mu^{\epsilon E}}$ are stiffness, strain to (electric, 
magnetic) field coupling constants (or piezo-electric and -magnetic coefficients), permittivity (dielectric) 
and (magnetic) permeability respectively. Considering the standard homogenization
procedure, the material functions $\mathbf{C^{EH}}$, $\mathbf{e}$, $\mathbf{e^M}$,
$\mathbf{\kappa^{\epsilon H}}$ and $\mathbf{\mu^{\epsilon E}}$,  are considered to 
be $\mathbf{Y}$-periodic functions in the 
unit cell domain defined as $\mathbf{Y}=[0,Y_1]\times[0,Y_2]\times[0,Y_3]$  
\citep{Jaya2014}. 
The functions involved in this expansion are assumed to be dependent on these two variables,
where one (i.e., $\mathbf{x}$) describing the "global" or average response of the structure
and the other (i.e., $\mathbf{y}$) describing the "local" or microstructural behaviour \cite{sanchez1980}. 
Here the two sets of variables $\mathbf{x}$ and $\mathbf{x}/\varepsilon$ take into 
account the two scales of the homogenization; the $\mathbf{x}$ variable is 
the macroscopic variable, whereas the $\mathbf{x}/\varepsilon$ variable takes into 
account the \emph{microscopic} geometry.

The detailed theoretical analysis is given elsewhere in a previous publication \citep{Jaya2014}. 
The local strain $\epsilon_{ij}^\varepsilon(\mathbf{x})$, electric field 
$E_j^\varepsilon(\mathbf{x})$ and the magnetic field 
$H_j^\varepsilon(\mathbf{x})$ too can be obtained once
the homogenized macroscopic problem is solved.
\begin{eqnarray}
\label{eq:S12}
\epsilon_{ij}^\varepsilon(\mathbf{x})&=&\epsilon_{ij}^0(\mathbf{x})+
\frac{\partial\eta^{mn}(\mathbf{x,y})}{\partial y_j}\nonumber \\
&~&\times \epsilon_{mn}(u^0(\mathbf{x}))
+\frac{\partial R^m(\mathbf{x,y})}{\partial y_j} 
\frac{\partial\varphi^0(\mathbf {x})}{\partial x_m}\nonumber \\
&~&+\frac{\partial\Psi^m(\mathbf{x,y})}{\partial y_j}\frac{\partial\psi^0
	(\mathbf {x})}{\partial x_m}
\end{eqnarray}
\begin{eqnarray}
\label{eq:S13}
E_j^\varepsilon(\mathbf{x})&=&-\big[\frac{\partial\varphi^0(\mathbf {x})}{\partial x_j}+\frac{\partial\eta^{mn}(\mathbf{x,y})}{\partial y_j} \nonumber\\
&~&\times\epsilon_{mn}(u^0(\mathbf{x})) 
+\frac{\partial R^m(\mathbf{x,y})}{\partial y_j}\frac{\partial\varphi^0(\mathbf {x})}
{\partial x_m} \nonumber\\
&~&+\frac{\partial \Psi^m(\mathbf{x,y})}{\partial y_j}
\frac{\partial\psi^0(\mathbf {x})}{\partial x_m}\big]
\end{eqnarray} 
\begin{eqnarray}
\label{eq:S14}
H_j^\varepsilon(\mathbf{x})&=&-\big[\frac{\partial\psi^0(\mathbf {x})}{\partial x_j}
+\frac{\partial\lambda^{mn}(\mathbf{x,y})}{\partial y_j} \nonumber\\
&~&\times{\epsilon_{mn}(u^0(\mathbf{x}))}
+\frac{\partial\Theta^m(\mathbf{x,y})}{\partial y_j} 
\frac{\partial\varphi^0(\mathbf {x})}{\partial x_m}\nonumber\\
&~&+\frac{\partial Q^m(\mathbf{x,y})}{\partial y_j}
\frac{\partial\psi^0(\mathbf {x})}{\partial x_m}\big]
\end{eqnarray}
The microscopic stress $\sigma_{ij}^\varepsilon(\mathbf{x})$, 
electrical displacement  $D_i^\varepsilon(\mathbf{x})$ and magnetic flux densities
$B_i^\varepsilon(\mathbf{x})$
at each point of the domain can be computed using the constitutive equations 
(\ref{eq:S8})--(\ref{eq:S10}), and the 
field equations as
\begin{eqnarray}
\label{eq:S15}
\sigma_{ij}^\varepsilon(\mathbf{x})&=&C_{ijkl}^\varepsilon(\mathbf{x})
\big(\frac{\partial{u_k^0(\mathbf{x})}}{\partial{x_l}}+
\frac{\partial u^1_k(\mathbf{x,y})}{\partial y_l}\big)\nonumber\\
&~&-e_{kij}^\varepsilon(\mathbf{x})\big(-\frac{\partial\varphi^0(\mathbf {x})}
{\partial x_k}-\frac{\partial\varphi^1(\mathbf {x,y})}{\partial y_k}\big)\nonumber\\
&~&-e_{kij}^{M~\varepsilon}(\mathbf{x})\big(-\frac{\partial\psi^0(\mathbf {x})}
{\partial x_k}\nonumber\\&&-\frac{\partial\psi^1(\mathbf {x,y})}{\partial y_k}\big)
\end{eqnarray}
\begin{eqnarray}
\label{eq:S16}
D_{i}^\varepsilon(\mathbf{x})&=& e_{ijk}^\varepsilon(\mathbf{x})
\big(\frac{\partial{u_j^0(\mathbf{x})}}{\partial{x_k}}+
\frac{\partial u^1_j(\mathbf{x,y})}{\partial y_k}\big)\nonumber\\
&~&+\kappa_{ij}^{\epsilon H~\varepsilon}(\mathbf{x})\big(-\frac{\partial\varphi^0(\mathbf {x})}
{\partial x_j}-\frac{\partial\varphi^1(\mathbf {x,y})}{\partial y_j}\big)\nonumber\\
&~&+\alpha_{ij}^\varepsilon(\mathbf{x})\big(-\frac{\partial\psi^0(\mathbf {x})}
{\partial x_j}-\frac{\partial\psi^1(\mathbf {x,y})}{\partial y_j}\big)
\end{eqnarray}
\begin{eqnarray}
\label{eq:S17}
B_{i}^\varepsilon(\mathbf{x})&=& e_{ijk}^{M~\varepsilon}(\mathbf{x})
\big(\frac{\partial{u_j^0(\mathbf{x})}}{\partial{x_k}}+
\frac{\partial u^1_j(\mathbf{x,y})}{\partial y_k}\big)\nonumber\\
&~&+\alpha_{ji}^\varepsilon(\mathbf{x})\big(-\frac{\partial\varphi^0(\mathbf {x})}
{\partial x_j}-\frac{\partial\varphi^1(\mathbf {x,y})}{\partial y_j}\big)\nonumber\\
&~&+\mu^{\epsilon E~\varepsilon}_{ij}(\mathbf{x})\big(-\frac{\partial\psi^0(\mathbf {x})}
{\partial x_j}\nonumber\\&&-\frac{\partial\psi^1(\mathbf {x,y})}{\partial y_j}\big)
\end{eqnarray}

The average fields are computed to be
\begin{eqnarray}
\label{eq:S18}
\langle\sigma_{ij}\rangle&=& \widetilde{C}^{EH}_{ijkl}(\frac{\partial u^0_k(\mathbf {x})}
{\partial x_l})
+\widetilde {e}_{kij}(\frac{\partial \varphi^0(\mathbf {x})}
{\partial x_k})\nonumber\\&&+\widetilde e^{~M}_{kij} (\frac{\partial \psi^0(\mathbf {x})}
{\partial x_k})
\\
\label{eq:S19}
\langle D_{i}\rangle &=&\widetilde {e}_{ijk}(\frac{\partial u^0_j(\mathbf {x})}
{\partial x_k})-
\widetilde {\kappa}^{\epsilon H}_{ij}(\frac{\partial \varphi^0(\mathbf {x})}
{\partial x_j})\nonumber\\&&-\widetilde \alpha_{ij}(\frac{\partial \psi^0(\mathbf {x})}
{\partial x_j})
\\
\label{eq:S20}
\langle B_{i}\rangle &=& \widetilde e^{~M}_{ijk}(\frac{\partial u^0_j(\mathbf {x})}
{\partial x_k})-
\widetilde {\alpha}_{ji}(\frac{\partial \varphi^0(\mathbf {x})}
{\partial x_j})\nonumber\\&&-\widetilde {\mu}^{\epsilon E}_{ij}
(\frac{\partial \psi^0(\mathbf {x})}
{\partial x_j})
\end{eqnarray}
The above macroscopic equations (i.e. they do not contain $\mathrm{y}$) can be computed
once the homogenized solution for $\mathbf{u}^0$, $\varphi^0$ and $\psi^0$ and that of the 
corresponding fields \emph{viz.,} $\frac{\partial u^0_j(\mathbf {x})}{\partial x_k}$, 
$\frac{\partial \varphi^0(\mathbf {x})}{\partial x_j}$ and 
$\frac{\partial \psi^0(\mathbf {x})} {\partial x_j}$ are prescribed. 
\section{Numerical Implementation}

We have developed a programming platform called POSTMAT (\emph{material postprocessing}) in this 
work for the solution of  the microscopic 
system of equations resulting form homogenization and the details are given 
elsewhere \citep{Jaya2014, Jayachandran-ScRep-2020}. A three-dimensional 
(3D) multiferroic finite element is conceived with five degrees of freedom (DOF)-three
DOFs for spatial displacements and one each for electric and magnetic potentials. 
Eight-noded isoparametric elements with $2\times2\times2$ Gauss-point integration 
are used to obtain solutions. Altogether there were nine microscopic equations that
should be solved for as much number of unknowns namely the characteristic functions
$\chi^{mn}_{i}, \eta^{mn}, \lambda^{mn}, R^{m},\Phi^{m}_{i}, \Theta^{m},
Q^{m}, \Psi^{m}$ and $\Gamma^{m}_k$, where the indices $\mathrm{m,n =1,2,3}$ \citep{Jaya2014}.
The problem is reduced to standard variational FEM, after the usual approximations of 
finite element formulation and can be expressed concisely as 
$\bm {Ku}=\bm{f}$
where $\bm K$ is the global stiffness matrix, $\bm u$ is the 
vector of unknown functions and $\bm f$ is the load vector.
Magnetoelectric multiferroic material, in general, can be considered as an 
aggregate of single crystalline crystallites/grains and hence the system altogether 
would be polycrystalline  nature. The unit cell or the representative volume element 
(RVE) conceived in this work is a volume containing a sufficiently large number of 
crystallites or grains that its properties can be considered as equivalent to that of 
the macroscopic sample. The electric polarization $\mathbf{P}$ as well as the 
magnetization $\mathbf{M}$ interspersed inside the crystallites could be
mapped using using some coordinate system. In a sense the underlying crystal 
orientation can encompass the orientations of  $\mathbf{P}$ or $\mathbf{M}$.
Thus we introduce the Euler angles $(\varphi,\theta,\psi)$ to quantify the 
crystal orientations of a multiferroic polycrystal (Fig.~S\ref{fig:S1}),
as the crystallites in an as-grown
sample are randomly oriented in the lattice space and
hence require three angles to describe its orientation with
reference to a fixed coordinate system. 
Here we use the so called
\emph{x-convention}, where the first and third rotation is through the y-axis 
(here it is $y_2^\prime$--axis)and the second rotation is through the intermediate 
x-axis (here it is  $y_1^\prime$--axis). Thus all the  physical
quantities $\mathbf{\gamma}^\prime_{ijklmn\dots}(\mathbf{y^\prime})$ 
expressed in a crystallographic coordinate system $\mathbf{y^\prime}$ would be 
coordinate-transformed to the local coordinate system $\mathbf{y}$ according
to the following scheme
\begin{equation}
\label{eq:S22}
\mathbf{\gamma}_{ijkl\dots}(\mathbf{y})=
e_{im}e_{jn}e_{kp}e_{lq}\dots\widetilde{\mathbf{\gamma}^\prime}_{mnpq\dots}
(\mathbf{y^\prime})
\end{equation}
before it is introduced for homogenization. (i.e., the FE and FM materials' 
electromechanical property data entered into the homogenization program are 
obtained with respect to the crystallographic
coordinates.) Here $e_{\mu\nu}$ are the Euler transformation
matrices \citep{Goldstein78}. 

For FEM simulation of the homogenization, a multiferroic crystallite is 
represented by a finite element in the microstructure. Thus we have a polycrystalline 
unite cell having as much number of crystallites as the number of finite elements
by which it is discretized. As-grown FE (or for that matter FM) polycrystal, 
often ends up in a near complete compensation of polarization (or magnetization) 
and the material consequently exhibit very small, if any, electric (or magnetic) 
effect until they are poled by the application of an electric (magnetic) field.
The orientation distribution of the crystallites (grains) in such a polycrystalline
material would be uniform with a standard deviation $\sigma\rightarrow \infty$ before 
poling (application of electric/magnetic field) and that after poling would
best be represented by a  distribution function with $\sigma\rightarrow 0$. 
Thus, any pragmatic configuration of orientation distribution of grains in 
multiferroic material would fit in a Gaussian distribution defined by the
probability distribution function  
The convergence of magnetoelectric properties with 
unit cell size allows us to determine the simulation-space independent, 
equivalent magnetoelectric properties of the composite. Convergence analyses, 
on magnetoelectric composites reveal that accuracy one derives from descretizing
the unit cell (in other words sampling of the unit cells with more number of 
grains or less number)  is minimal above 1000 elements (grains) 
\citep{Jayachandran-CMS-2018}. Consequently, we kept unit cells' sizes 
above 1000 finite elements in this study.  

\section{Results}
Piezomagnetism drives materials to acquire 
magnetization upon application of mechanical stress. Since in BaTiO$_3$--CoFe$_2$O$_4$ 
composite system this phenomenon plays a crucial role in the overall magnetization 
that it acquires consequent to the electrical loading.
 The distribution of local (\emph{microscopic}) fields \emph{viz}., the stress $\sigma^\varepsilon$, 
 von Mises stress $\sigma_v^\varepsilon$ and strain $\epsilon^\varepsilon$ consequent to 
 the application of an external electric field 
 along the $y_3$--axis of the composite microstructure is 
 displayed in Fig.~S\ref{fig:S1}. The histograms in Fig.~S\ref{fig:S2} show the frequency of 
 stress and strains that spread across the microstructure. 
   